\documentstyle[12pt]{article}
\begin{document}
\begin{center}
{\Large {\bf $\pi K$ scattering in effective 
chiral theory of mesons}}\\[5mm]
Bing An Li\\
{\small Department of Physics and Astronomy, University of Kentucky\\
Lexington, Kentucky 40506, USA}\\[2mm]
Dao-Neng Gao\\
{\small Center for Fundamental Physics, University of Science and
Technology of China}\\
{\small Hefei, Anhui 230026, People's Republic of China}\\[2mm]
Mu-Lin Yan\\
{\small Chinese Center for Advanced Science and Technology
(World Lab)\\
P.O.Box 8730, Beijing, 100080, People's Republic of China}\\
{\small and}\\
{\small Center for Fundamental Physics and Center for Nonlinear
Science\\
University of Science and Technology of China,
Hefei, Anhui 230026\\
 People's Republic of China}
\end{center}
\vspace{8mm}
\begin{abstract}
\noindent
In the framework of U(3)$_L\times$ U(3)$_R$ chiral theory of mesons,
$\pi K$
elastic scattering is studied. The s-wave and p-wave scattering
lengths
($a^{3/2}_0$, $a^{1/2}_0$, $a^{3/2}_1$, and $a^{1/2}_1$) and
phase shifts
($\delta^{3/2}_0$, $\delta^{1/2}_0$, $\delta^{3/2}_1$, and
$\delta^{1/2}_1$),
the total elastic scattering cross section  $\sigma_{\rm tot}
(\pi^+ K^+)$
and $\sigma_{\rm tot}(\pi^- K^+)$, and the p-wave cross section
$\sigma_p(\pi^- K^+)$ are calculated.
Theoretical results are in agreement with the
experimental data.
There is no new parameter in this study.
\end{abstract}
 
\vspace{4mm}
 
\newpage
 
The chiral SU(3)$_L\times$ SU(3)$_R$ symmetry of quantum
chromodynamics (QCD)
Lagrangian in the limit of $m_u=m_d=m_s=0$
is spontaneously
broken to SU(3)$_V$ with the appearances of eight Goldstone
pseudoscalar-particles ($\pi, K, \eta$). At low energies,
these Goldstone
Bosons play very important role in
chiral dynamics of the strong interaction.
It is well known that the reactions
$ \pi \pi\longrightarrow \pi \pi$ and $\pi K\longrightarrow
\pi K$ scattering are strongly related to chiral dynamics.
The investigation of $\pi \pi$ scattering has a long history
\cite{We, Gr, DHT, GL1, DRV, BKM1, Li1} and it tests all kinds
of chiral theory of meson physics.
$\pi K$ scattering is another test of chiral dynamics.
Chiral symmetry has been exploited in the study of $\pi$ K
scattering\cite{BKM1, BKM2, DP, BT, BS}.
In Ref. \cite{BKM2,BS}
chiral perturbation
theory (ChPT) has been applied to $\pi$K scattering.
In Ref.\cite{BKM1} resonances are incorporated.
 
In Refs.\cite{Li1,Li2}
a chiral theory of mesons including pseudoscalar,
vector, and
axial-vector mesons (in brief  U(3)$_L\times$U(3)$_R$ theory
below) has been proposed.
This theory is phenomenologically successful
\cite{Li3, Li4, GLY, WY, Li5}. It has been applied to
study $\pi \pi$ scattering\cite{Li1}. Theoretical results
agree well with data. In terms of $\pi$K scattering three coefficients
of the chiral perturbation theory have been determined[17]. 
In this paper we investigate
$\pi K$ scattering in the framework of this effective chiral
theory of mesons.
 
The three invariant quantities of
$\pi(p_1)+K(p_2)\longrightarrow \pi(p_3)+K(p_4)$ are defined as
\begin{eqnarray*}
& &s=(p_1+p_2)^2=(p_3+p_4)^2,\\
& &t=(p_1-p_3)^2=(p_2-p_4)^2,\\
& &u=(p_1-p_4)^2=(p_2-p_3)^2,
\end{eqnarray*}
with $s+t+u=2(m_\pi^2+m_K^2)$.
$T^{3/2}(s, t, u)$ and $T^{1/2}(s, t, u)$ are the
two independent isospin amplitudes of $\pi$K scattering
\begin{eqnarray}
& &T^{3/2}(s, t, u)=T_{\pi^+ K^+}(s, t, u),\\
& &T^{1/2}(s, t, u)=\frac{3}{2}T_{\pi^- K^+}(s, t, u)-
   \frac{1}{2}T^{3/2}(s, t, u).
\end{eqnarray}
 
The partial wave amplitudes are defined as
\begin{equation}
T^I_l(s)=\frac{1}{32\pi}\int^1_{-1}d\hspace{0.03in}{\rm cos}{\theta}
P_l({\rm cos}{\theta}) T^I(s, t, u).
\end{equation}
In terms of
real phase shifts $\delta^I_l(s)$, $T^{I}_{l}(s)$ is
expressed as
\begin{eqnarray}
& &T^I_l(s)=\frac{\sqrt{s}}{2 q}\frac{1}{2 i}(e^{2i \delta^I_l(s)}-1),
\nonumber
\\
& & q=\frac{1}{2\sqrt{s}}\sqrt{[s-(m_K+m_\pi)^2][s-(m_K-m_\pi)^2]}.
\end{eqnarray}
Here $q$ is the pion's or kaon's momentum and $\theta$ is the
scattering angle in the frame of center of mass.
 
At low energies (small $q$), the $\pi K$ scattering
length $a^I_l$ is defined
\begin{equation}
\mbox{Re}\hspace{0.05in}T^I_l(s)=\frac{\sqrt{s}}{2}
q^{2l}(a^I_l+O(q^2)).
\end{equation}
 
All the meson vertices involved in $\pi$K scattering are derived
from the real part of the effective Lagrangian
(eq.(3) in Ref.\cite{Li2}). The vertices
${\cal L}^{\pi\pi KK}$, ${\cal L}^{\pi\pi\rho}$, ${\cal L}^{\rho K
K}$,
and ${\cal L}^{\pi KK^{*}}$ are involved in $\pi$K scattering
\begin{eqnarray}
&&{\cal L}^{\pi\pi KK}=
{1\over f^{2}_{\pi}}(1-{6c\over g})\pi_{i}\partial_{\mu}\pi_{i}
K_{a}\partial^{\mu}K_{a}-{1\over 6 f_\pi^2}(1-{6c\over
g})(m_\pi^2+m_K^2)\pi_i\pi_i K_a K_a\nonumber\\
&&+{1\over f^{4}_{\pi}}\{{1\over\pi^{2}}
(1-{2c\over g})
^{2}{4c^{2}\over g^{2}}-{8c^{4}\over g^{2}}\}\partial_{\mu}\pi_{i}
\partial^{\mu}\pi_{i}\partial_{\nu}K_{a}\partial^{\nu}K_{a}
\nonumber \\
&&+{1\over f^{4}_{\pi}}\{{1\over\pi^{2}}(1-{2c\over g})^{2}(1-{
4c\over g})+{8c^{4}\over g^{2}}\}\partial_{\mu}\pi_{i}\partial_{\nu}
\pi_{i}\partial^{\mu}K_{a}\partial^{\nu}K_{a}\nonumber \\
&&-{2\over f^{2}_{\pi}}(1-{6c\over g})\epsilon_{ijk}f_{abk}\partial_
{\mu}\pi_{i}\pi_{j}\partial^{\mu}K_{a}K_{b}\nonumber \\
&&+{1\over f^{4}_{\pi}}\{{1\over\pi^{2}}(1-{2c\over g})^{2}
({16c^{2}\over g^{2}}+{8c\over g}-2)-{48c^{4}\over g^{2}}\}
\epsilon_{ijk}f_{abk}\partial_{\mu}\pi_{i}\partial_{\nu}\pi_{j}
\partial^{\mu}K_{a}\partial^{\nu}K_{b},  \\
&&{\cal L}^{\rho\pi\pi}=f(q^{2})\epsilon_{ijk}\rho^{i}_{\mu}
\pi_{j}\partial^{\mu}\pi_{k},\\
&&{\cal L}^{\rho KK}=f(q^{2})f_{abi}\rho^{i}_{\mu}
K_{a}\partial^{\mu}K_{b},\\
&&{\cal L}^{K^{*}K\pi}=f(q^{2})f_{abi}K^{a}_{\mu}
(\partial^{\mu}\pi_{i}K_{b}-\pi_{i}\partial^{\mu}K_{b}),
\end{eqnarray}
where
\begin{equation}
f(q^{2})={2\over g}(1+\beta q^{2}),\;\;\; \beta={1\over2\pi^{2}f^{2}_{\pi}}
\{(1-{2c\over g})^{2}-4\pi^{2}c^{2}\}.
\end{equation}
 
The amplitudes of this process from the contact terms
are derived from ${\cal L}^{\pi\pi KK}$(with the index $D$)
\begin{eqnarray}
& &T^{{3\over2}}(u,t,s)_D=T_{\pi^- K^+}(s, t, u)_D,\nonumber \\
& &T^{{3\over2}}(s, t, u)_D=\frac{2}{f_\pi^2}(1-\frac{6c}{g})(m_\pi^2
+m_K^2-s)-
\frac{2}{3f_\pi^2}(1-\frac{6c}{g})(m_\pi^2+m_K^2)\nonumber\\
& &+[\frac{1}{\pi^2 f_\pi^4}(1-\frac{2c}{g})^2\frac{4c^2}{g^2}-
\frac{8c^4}{f_\pi^4 g^2}](t-2m_\pi^2)(t-2m_K^2)\nonumber\\
& &+[\frac{1}{\pi^2 f_\pi^4}(1-\frac{2c}{g})^2(1-\frac{4c}{g}
-\frac{4c^2}{g^2})+
\frac{16 c^4}{f_\pi^4 g^2}](s-m_\pi^2-m_K^2)^2\nonumber \\
& &+[\frac{1}{\pi^2 f_\pi^4}(1-\frac{2c}{g})^2\frac{4c^2}{g^2}-
\frac{8c^4}{f_\pi^4 g^2}](u-m_\pi^2-m_K^2)^2,\\
& &T^{{1\over2}}(s,t,u)_D=\frac{3}{2}T_{\pi^- K^+}(s,t,u)_D-
\frac{1}{2} T^{3/2}(s, t, u)_D,\nonumber \\
& &T^{{1\over2}}(s,t,u)_D={1\over f^2_{\pi}}(1-{6c\over g})(2m^{2}_{\pi}
+2m^{2}_{K}+s-3u)-\frac{2}{3f_\pi^2}(1-\frac{6c}{g})(m_\pi^2+m_K^2)\nonumber
\\
& &+{1\over f^{4}_{\pi}}\{{1\over \pi^{2}}(1-{2c\over g})^{2}{4c^{2}
\over g^{2}}-{8c^{4}\over g^{2}}\}(t-2m^{2}_{\pi})(t-2m^{2}_{K})
\nonumber \\
& &+{1\over f^{4}_{\pi}}\{{1\over\pi^{2}}(1-{2c\over g})^{2}
(\frac{3}{2}-{6c\over g}-{8c^{2}\over g^{2}})+{28c^{4}\over g^{2}}
\}(u-m_\pi^2-m_K^2)^2\nonumber \\
& &+{1\over f^{4}_{\pi}}\{{1\over\pi^{2}}(1-{2c\over g})^{2}
({8c^{2}\over g^{2}}-\frac{2c}{g}-\frac{1}{2})-{20c^{4}\over
g^{2}}
\}(s-m^{2}_{\pi}-m^{2}_{K})^{2}.
\end{eqnarray}
Where $f_\pi$ is the decay constant of the pion,  $g$ is the universal
coupling
constant of the $U(3)_{L}\times U(3)_{R}$
theory, and it has been fixed $g$=0.35 in
Refs. \cite{Li1, Li2}.
Following equations have been used in deriving
eqs.(11,12),
\begin{eqnarray}
& &\frac{F^2}{f_\pi^2}(1-\frac{2c}{g})=1,\nonumber\\
& &c=\frac{f_\pi^2}{2g m_\rho^2}.
\end{eqnarray}
 
The resonance parts of the two amplitudes are obtained from the
vertices(7,8,9)
\begin{eqnarray}
T^{{3\over2}}(s, t, u)_R&=&\frac{1}{2}f^2(t) \frac{1}{t-m_\rho^2}
(s-u)\nonumber\\
& &+\frac{1}{2}f^2(u) \frac{1}{u-m_{K^*}^2}-s-t-
\frac{(m_\pi^2-m_K^2)^2}{m_{K^*}^2}],\\
T^{{1\over2}}(s, t, u)_R&=&\frac{3}{2} T_{\pi^- K^+}(s, t, u)_R-
\frac{1}{2} T^{3/2}(s, t, u)_R\nonumber\\
&=&f^2(t) \frac{1}{t-m_\rho^2}(u-s) \nonumber \\
& &+\frac{3}{4}f^2(s)\frac{1}{s-m_{K^*}^2+i\sqrt{s}
\Gamma(s)}[u-t-
\frac{(m_\pi^2-m_K^2)^2}{m_{K^*}^2}- \nonumber \\
& &-\frac{1}{4}f^2(u)
\frac{1}{u-m_{K^*}^2}[s-t-\frac{(m_\pi^2-m_K^2)^2}{m_{K^*}^2}],
\end{eqnarray}
where $\Gamma(s)$ is the decay width of $K^{*}$
\begin{equation}
\Gamma(s)=f^2(q^2)\frac{q^3}{8\pi s},
\end{equation}
$q$ is defined in eq.(4).
It is necessary to point out that in the amplitudes(11,12,14,15) all the
parameters have been fixed in Refs.[7,12] and there is no new
parameter.
 
The scattering lengths can be derived from the amplitudes(11,12,14,15).
However, the leading terms of the s-wave scattering lengths
are proportional to either $m_{\pi}$ or $m_{K}$. The quark mass
term $\bar{\psi}M\psi$(M is the quark matrix) contributes
to the s-wave scattering lengths too. The effective Lagrangian
with quark mass term has been derived in Ref.[17]. Using the
formalism developed in Ref.[17], to the leading order in quark masses
the contribution of the quark
mass term to $\pi$K scattering can be found from
\begin{equation}
{\cal L}={1\over2}\int\frac{d^{D}p}{(2\pi)^{D}}\frac{m}
{p^{2}+m^{2}}Tr(\hat{u}M+Mu),
\end{equation}
where \(u=exp\{i\gamma_{5}{2\over f_{\pi}}
(\tau_{i}\pi_{i}+\lambda_{a}K_{a})\}\) and
\(\hat{u}=exp\{-i\gamma_{5}{2\over f_{\pi}}
(\tau_{i}\pi_{i}+\lambda_{a}K_{a})\}\), m is a parameter defined in
\cite{Li1}. Eq.(7) leads to
\begin{equation}
{\cal L}={1\over 3f^{2}_{\pi}}<\bar{\psi}\psi>\{{3\over2}
(m_{u}+m_{d})+m_{s}\}(2\pi^{+}\pi^{-}+\pi^{0}\pi^{0})
(K^{+}K^{-}+K^{0}\bar{K}^{0}),
\end{equation}
$<\bar{\psi}\psi>$ is the quark condensate and defined in Ref.[17]
\begin{equation}
<\bar{\psi}\psi>=-\frac{mDN_{c}}{(2\pi)^{D}}\int d^{D}p
\frac{1}{p^{2}+m^{2}},
\end{equation}
where $N_{c}$ is the number of color.
To the leading order in quark masses pion and kaon masses are
expressed as[17]
\begin{eqnarray}
m^{2}_{\pi}&=&-{4\over f^{2}_{\pi}}<\bar{\psi}\psi>(m_{u}+m_{d}),
\nonumber \\
m^{2}_{K^{+}}&=&-{4\over f^{2}_{\pi}}<\bar{\psi}\psi>(m_{u}+m_{s}),
\nonumber \\
m^{2}_{K^{0}}&=&-{4\over f^{2}_{\pi}}<\bar{\psi}\psi>(m_{d}+m_{s}).
\end{eqnarray}
From Eqs.(18,20) we obtain
\begin{equation}
{\cal L}={1\over3f^{2}_{\pi}}\{m^{2}_{\pi}+{1\over2}(m^{2}_{K^{+}}
+m^{2}_{K^{0}})\}(2\pi^{+}\pi^{-}+\pi^{0}\pi^{0})(K^{+}K^{-}+K^{0}\bar{K}^{0}).
\end{equation}
 
Using eqs.(3) and (5), the s-wave and p-wave scattering
lengths from the contact parts of the amplitudes(11,12) and Eq.(21) are found
\begin{eqnarray}
{a^{3/2}_0}_D&=&\frac{1}{2\pi}[-\frac{m_\pi m_K}{f_\pi^2(m_\pi+m_K)}
(1-
\frac{6c}{g})+\frac{c}{f_\pi^2 g}\frac{m_\pi^2+m_K^2}{m_\pi+m_K}
\nonumber\\
&&+\frac{1}{\pi^2 f_\pi^4}(1-\frac{2c}{g})^4\frac{m_\pi^2 m_K^2}
{m_\pi+m_K}],\\
{a^{3/2}_1}_D&=&-\frac{1}{6\pi}[\frac{1}{\pi^2 f_\pi^4}
(1-\frac{2c}{g})^2
\frac{4c^2}{g^2}-\frac{8c^4}{f_\pi^4 g^2}]\frac{(m_K-m_\pi)^2}
{m_K+m_\pi},\\
{a^{1/2}_0}_D&=&\frac{1}{\pi f_\pi^2}[(1-\frac{6c}{g})
\frac{m_\pi m_K}{m_\pi+m_K}
+\frac{c}{2g}\frac{m_\pi^2+m_K^2}{m_\pi+m_K}]\nonumber\\
&&+\frac{1}{2\pi^3 f_\pi^4}(1-\frac{2c}{g})^2(1-\frac{4c}{g})
\frac{m_\pi^2 m_K^2}{m_\pi+m_K},\\
{a^{1/2}_1}_D&=&\frac{1}{4\pi f_\pi^2}(1-\frac{6c}{g})\frac{1}
{m_K+m_\pi}-
\frac{1}{6\pi}[\frac{1}{\pi^2 f_\pi^4}(1-\frac{2c}{g})^2
\frac{4c^2}{g^2}-
\frac{8c^4}{f_\pi^4 g^2}]\frac{m_\pi^2+m_K^2}{m_\pi+m_K}
\nonumber\\
&&+\frac{1}{3\pi}[\frac{1}{\pi^2 f_\pi^4}(1-\frac{2c}{g})^2
(\frac{3}{2}-
\frac{6c}{g}-\frac{8c^2}{g^2})+\frac{28c^4}{f_\pi^4 g^2}]
\frac{m_\pi m_K}{m_\pi+m_K}.
\end{eqnarray}
From the resonance parts of the amplitudes(14,15), we obtain
\begin{eqnarray}
{a^{3/2}_0}_R&=&-\frac{m_\pi m_K}{\pi g^2 m_\rho^2(m_\pi+m_K)}-
\frac{m_\pi+m_K}{4\pi g^2 m_{K^*}^2}[1+\beta(m_\pi-m_K)^2]^2,\\
{a^{3/2}_1}_R&=&-\frac{m_\rho^2+4 m_\pi m_K+8\beta m_\pi m_K
m_\rho^2}{6\pi g^2 m_\rho^4(m_\pi+m_K)}+
\frac{1+\beta^2(m_K^2-m_\pi^2)^2}{6\pi g^2 m_{K^*}^2(m_\pi+m_K)}
\nonumber\\
& &-\frac{\beta^2(m_\pi^2+m_K^2)}{3\pi g^2 (m_\pi+m_K)}+
\frac{(1+\beta m_{K^*}^2)^2(m_\pi^2+m_K^2)}{3\pi g^2 m_{K^*}^2
[m_{K^*}^2-
(m_K-m_\pi)^2](m_\pi+m_K)},\\
{a^{1/2}_0}_R&=&\frac{2 m_\pi m_K}{\pi g^2 m_\rho^2(m_\pi+m_K)}+
\frac{m_\pi+m_K}{8\pi g^2 m_{K^*}^2}[1+\beta(m_\pi-m_K)^2]^2
\nonumber\\
& &-\frac{3(m_K-m_\pi)^2[1+\beta(m_\pi+m_K)^2]^2}{8\pi g^2
m_{K^*}^2(m_\pi+m_K)},\\
{a^{1/2}_1}_R&=&\frac{m_\rho^2+4 m_\pi m_K+8\beta m_\pi m_K
m_\rho^2}{3\pi g^2 m_\rho^4(m_\pi+m_K)}+
\frac{[1+\beta(m_\pi+m_K)^2]^2}{2\pi g^2[m_{K^*}^2-(m_\pi+m_K)^2
](m_\pi+m_K)}\nonumber \\
& &-\frac{1+\beta^2(m_K^2-m_\pi^2)^2}{12\pi g^2 m_{K^*}^2
(m_\pi+m_K)}
+\frac{\beta^2 (m_\pi^2+m_K^2)}{6\pi g^2 (m_\pi+m_K)}\nonumber \\
& &-\frac{(1+\beta m_{K^*}^2)^2(m_\pi^2+m_K^2)}{6\pi g^2 m_{K^*}^2
[m_{K^*}^2-
(m_K-m_\pi)^2](m_\pi+m_K)}.
\end{eqnarray}
Numerical calculations show that
the vector resonance exchanges
are dominant in $\pi K$ scattering.
The numerical results of the scattering lengths
are listed in Table 1. The data can be traced back from
Ref. \cite{BKM1}.
$a^I_0$ (s-wave) and $a^I_1$ (p-wave) are given in
units of $m_\pi^{-1}$ and $m_\pi^{-3}$ respectively.
The theoretical predictions of $a^{1/2}_1$,
$a^{1/2}_0$, and $a^{3/2}_0$ are in reasonable
agreement with the experimental data. It is found that
in $a^{3/2}_1$ the $\rho$ and $K^*$ resonances have opposite
signs,
which leads to strong cancellation, so the value of $a^{3/2}_1$
is very small.
\begin{table}[t] \begin{center}
\begin{tabular}{r|c|c}
\hline
       &   theoretical results & experimental data \\ \hline
$a^{1/2}_0$ &  0.13  & 0.13 ... 0.24\\
$a^{3/2}_0$ &  --0.05 & --0.13 ... --0.05\\
$a^{1/2}_1$ &  0.017 & 0.017 ... 0.018\\
$a^{3/2}_1$ & --1.68$\times$10$^{-4}$& -----\\ \hline
\end{tabular}
\caption{The s-wave and p-wave $\pi K$ scattering lengths.}
\end{center}
\end{table}
 
From eq.(4),
the phase shifts of the
partial waves
of  $\pi K$ scattering are defined as
\begin{equation}
\delta^I_l(s)={\rm arctan}(\frac{\mbox{Im}\hspace{0.05in} T^I_l(s)}
{\mbox{Re}\hspace{0.05in} T^I_l(s)}).
\end{equation}
Using the unitarity constraint on the partial-wave amplitude
of
elastic scattering[6], at the leading order it is found
\begin{equation}
\delta^I_l(s)=\mbox{arctan}(\frac{2q}{\sqrt{s}} \mbox{Re}
\hspace{0.05in} T^I_l(s)).
\end{equation}
 
The isospin 1/2 p-wave phase shifts $\delta^{1/2}_1$ are shown in
Fig. 1 and Fig. 2.
$\delta^{1/2}_1$
is dominated by the $K^*(892)$ resonance.
The theoretical results are
in good agreement with the data.
The isospin 3/2 p-wave phase shifts are very
small(Fig. 3). This is consistent with the result
presented in Ref. \cite{BKM1}.
 
The s-wave phase shifts are shown in Fig. 4 and Fig. 5 respectively.
Theoretical predictions of $\delta^{1/2}_0$
and $\delta^{3/2}_0$ are in agreement with the data within the error
bars.
 
The differential cross section of
elastic $\pi$K scattering is expressed as
\begin{equation}
\frac{d\sigma}{d t}=\frac{1}{64\pi s q^2}|T|^2,
\end{equation}
The total cross
section of $\pi$K scattering is written as
\begin{equation}
\sigma_{\rm tot}=\frac{1}{32\pi s}\int^1_{-1} d\hspace{0.05in}
{\rm cos}\theta
|T|^2.
\end{equation}
 
Using Eqs.(11,12,14,15), the total cross sections are
calculated.
Theoretical results of
$\sigma_{\rm tot}(\pi^+ K^+)$ are in the range of 0.63 mb
$\sim$ 2.0 mb,
which agrees with the experimental value $\sigma_{\rm tot}^
{3/2}\sim$ 1.8 mb
given in Ref. \cite{Bing}.
 
The amplitude of $\pi^- K^+$ scattering has
I=1/2 and I=3/2 two components.
$\delta^{3/2}_1$
is very small (Fig. 3), so it can be ignored in the calculation
of the
p-wave cross section in terms of phase shifts.
Thus, we have
\begin{equation}
\sigma_p(\pi^- K^+)=\frac{16\pi}{3 q^2} {\rm sin}^2
\delta^{1/2}_1.
\end{equation}
In Fig. 7, the p-wave cross section of $\pi^- K^+$ elastic
scattering
is shown.
The
predictions of $\sigma_p(\pi^- K^+)$ are in good agreement with the
data.
 
Our calculation shows
that the contributions from s-wave and p-wave are
dominant
in the $\pi K$ elastic scattering at low energies, which is
consistent with one claimed in Ref. \cite{Matison}.
 
To conclude, in terms of the
U(3)$_L\times$ U(3)$_R$ theory of mesons a parameter free
study of
$\pi$K scattering is presented. Theoretical results of
the s-wave and p-wave
scattering lengths, the phase shifts, and the cross sections
are in good agreement with the data.
The vector resonances are dominant
in $\pi K$ scattering.
 
B.A.Li is partially supported by DOE Grant No. DE-91ER75661.
D.N.Gao and M.L.Yan are partially supported by NSF of China through
C.N.Yang.


\newpage
\leftline{\bf Caption}
\begin{description}
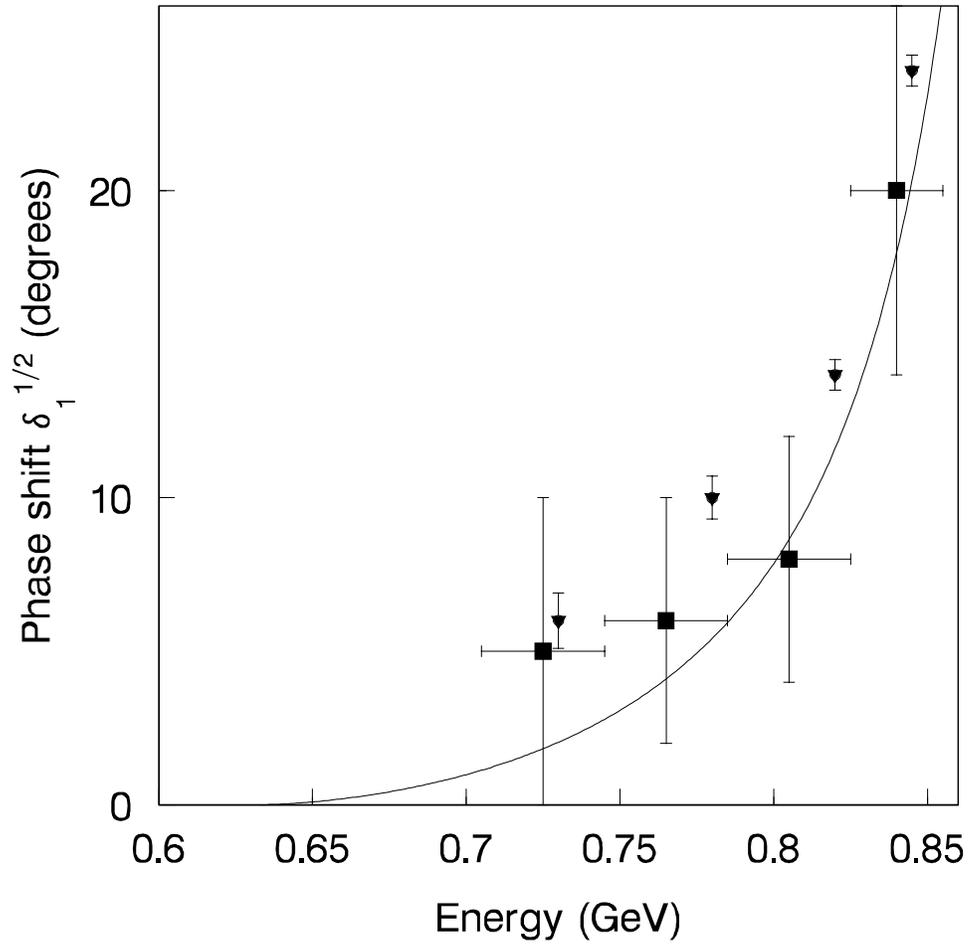
\item[FIG. 1] The I=1/2 p-wave $\delta^{1/2}_1(s)$ for $\sqrt{s}\le$ 0.86 GeV.
The solid line denotes the theoretical results.
The data are from Ref. \cite{Mercer} (solid squares) and
Ref. \cite{Esta} (solid inverted triangles).
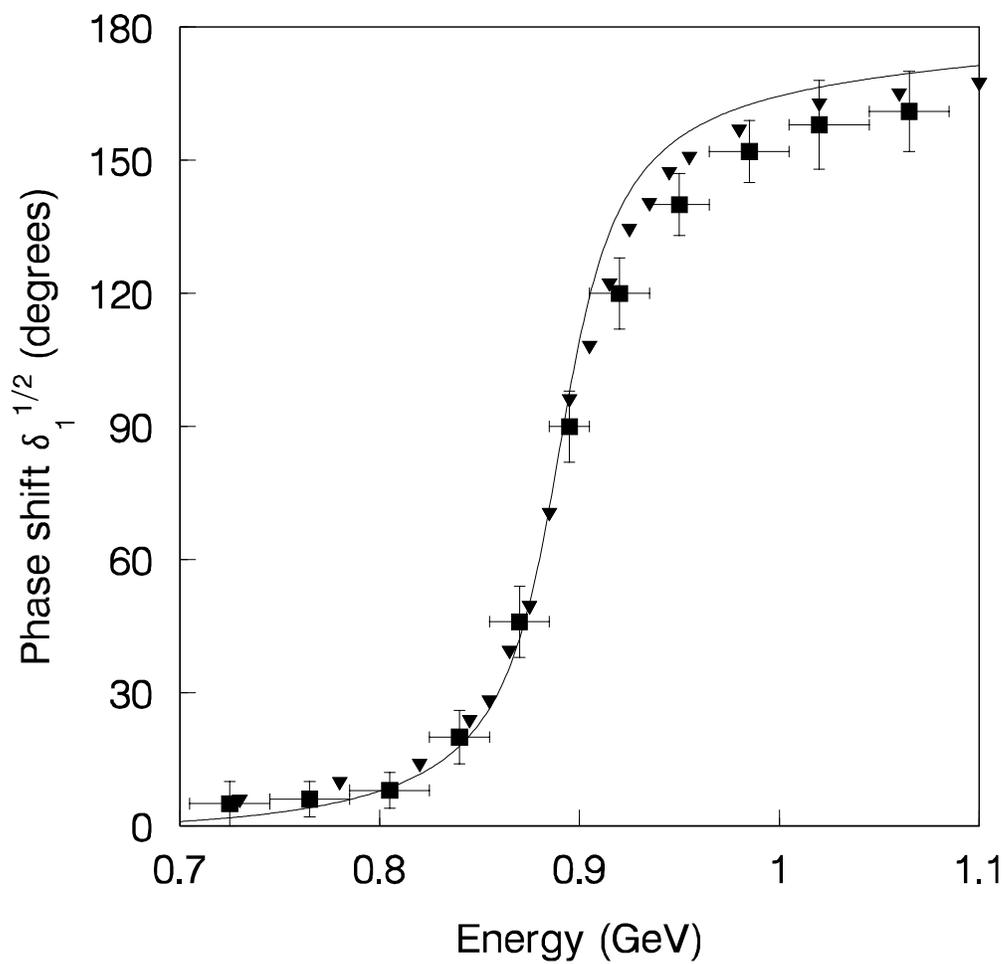
\item[FIG. 2] The I=1/2 p-wave $\delta^{1/2}_1(s)$ for
0.70 GeV$\le\sqrt{s}\le$ 1.1 GeV. For the notations, see Fig. 1.
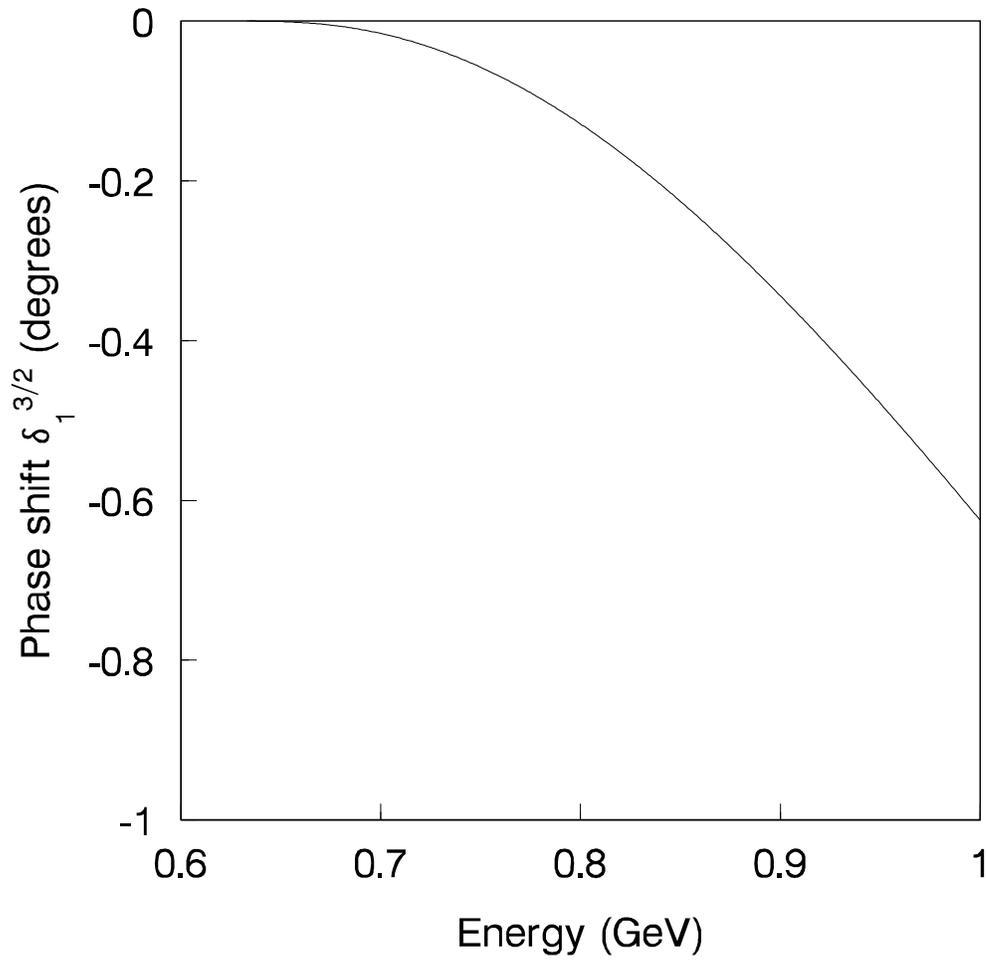
\item[FIG. 3] The I=3/2 p-wave $\delta^{3/2}_1(s)$ for $\sqrt{s}\le$ 1.0 GeV.
The solid line denotes the theoretical results.
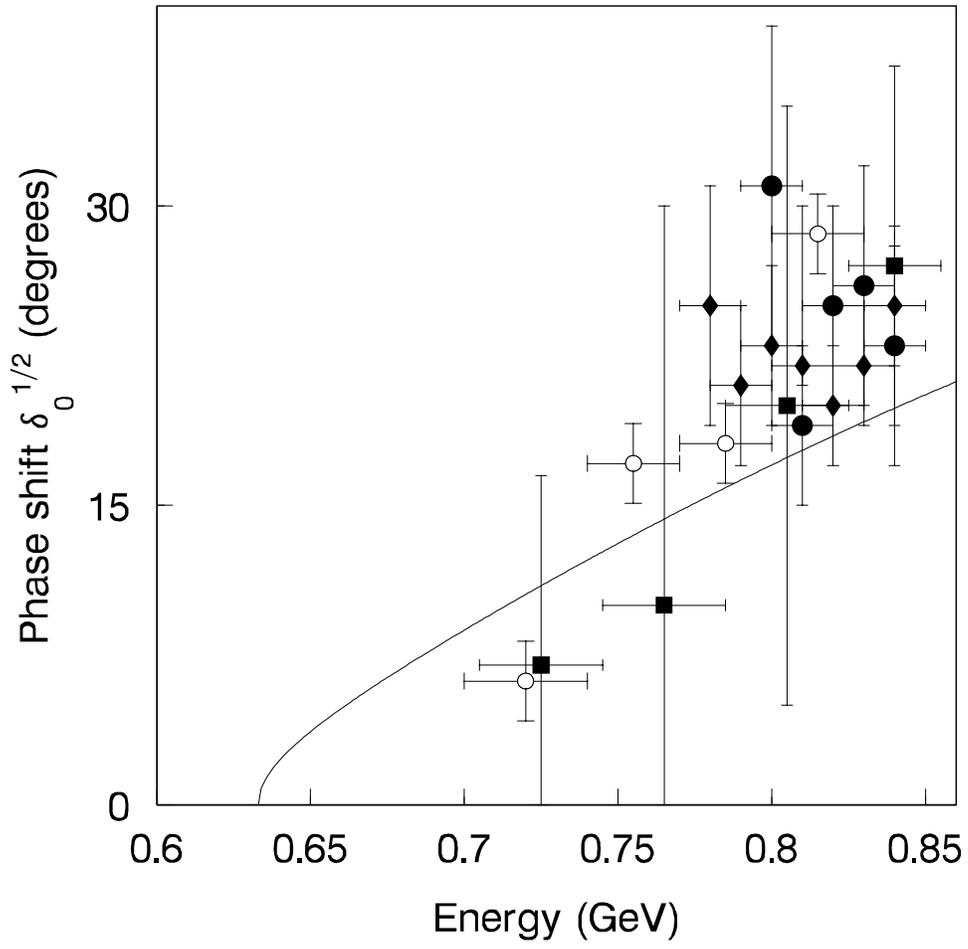
\item[FIG. 4] The I=1/2 s-wave $\delta^{1/2}_0(s)$ for $\sqrt{s}\le$ 0.86 GeV.
The solid line denotes the theoretical results.
The key to the data is as follows: Ref. \cite{Matison} (solid circles),
Ref. \cite{Bing} (solid diamonds),  Ref. \cite{Mercer} (solid squares),
Ref. \cite{Baker} (open circles).
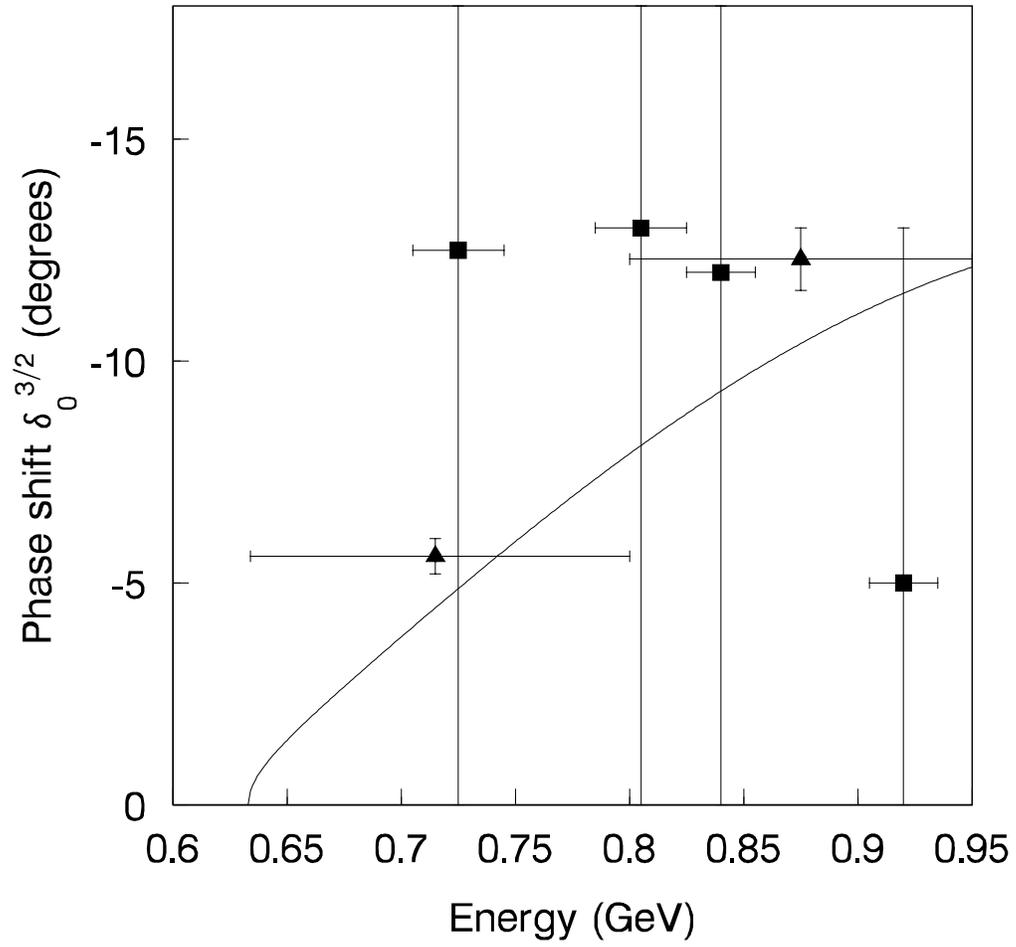
\item[FIG. 5] The I=3/2 s-wave $\delta^{3/2}_0(s)$ for $\sqrt{s}\le$ 0.95 GeV.
The solid line denotes the theoretical results.
The data are from Ref. \cite{Mercer} (solid squares) and
Ref. \cite{Jong} (solid triangles).
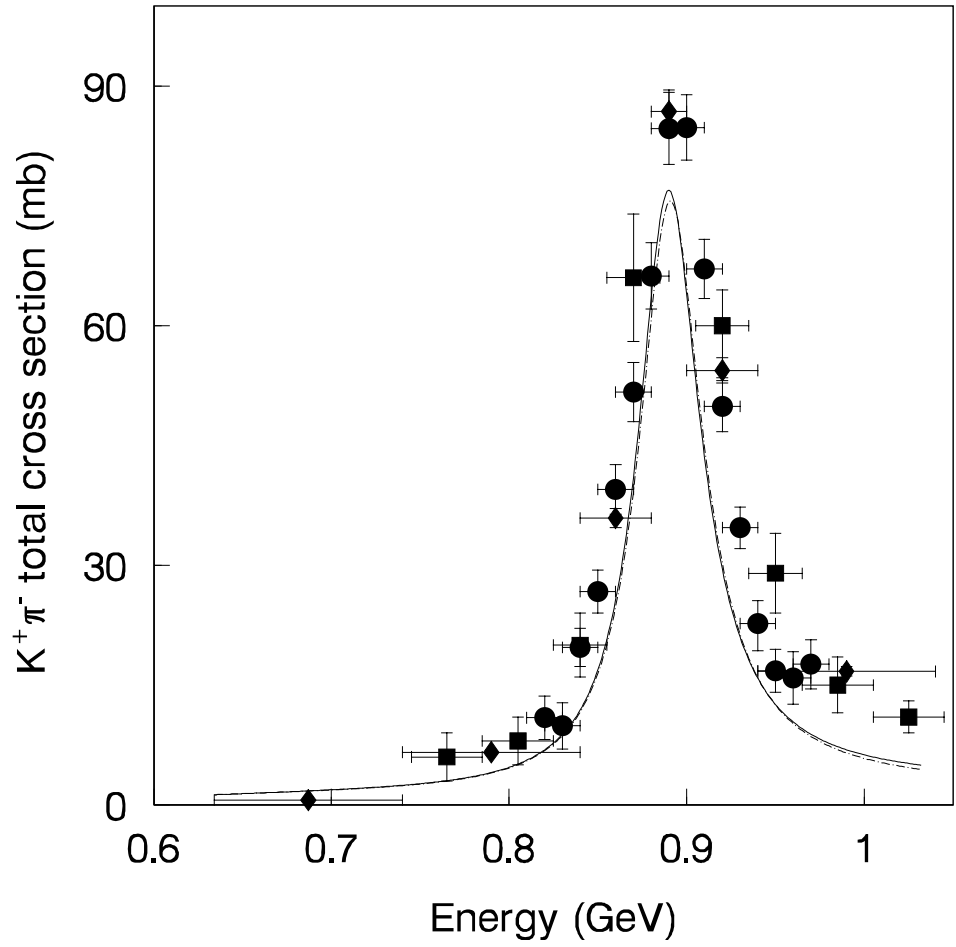
\item[FIG. 6] The total elastic cross section $\sigma_{\rm tot}(\pi^- K^+)$
for $\sqrt{s}\le$ 1.05 GeV. For the notations, see Fig. 4. The solid line is
obtained by using eq. (30), and the dot-dashed line is from the partial wave
method.
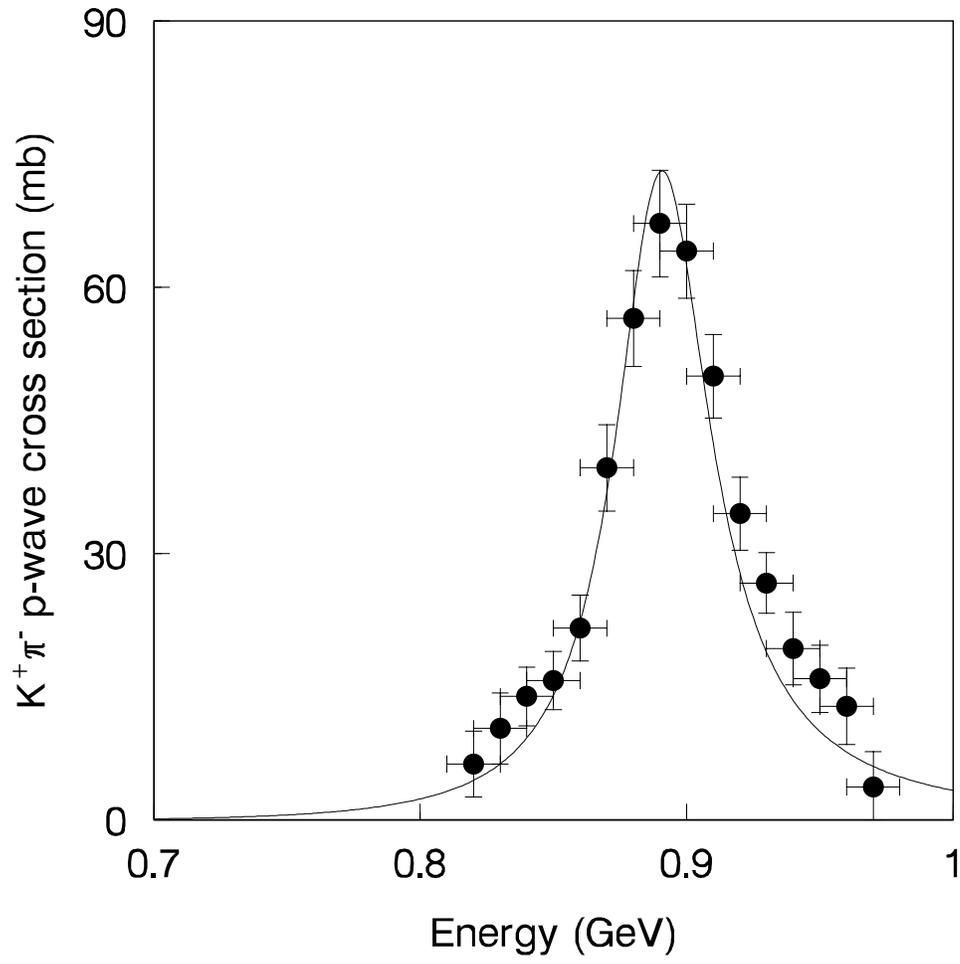
\item[FIG. 7] The p-wave elastic cross section $\sigma_p(\pi^- K^+)$
for $\sqrt{s}\le$ 1.0 GeV. The solid line denotes the theoretical results.
The data are from Ref. \cite{Matison} (solid circles).
\end{description}
\end{document}